\documentclass[twocolumn]{jpsj3}
\usepackage{txfonts}

\title{An enhancement mechanism of two-particle exchange interactions\\
in single- and multi-layer cuprate superconductors}

\author{Shingo Teranishi\thanks{teranishi@artemis.mp.es.osaka-u.ac.jp}, 
Satoaki Miyao, Kazutaka Nishiguchi, and Koichi Kusakabe}
\inst{
Depatment of Materials Engineering Science, Graduate School of Engineering Science, 
Osaka University, \\
1-3 Machikaneyama-cho, Toyonaka Osaka 560-8531, Japan}

\abst{
To explore material dependence of layered cuprate superconductors, 
we examine effective two-particle interactions for Hg1201 and Tl1201, 
where Tl1201 having a nearly half value of $T_c$ of 
Hg1201 even at the optimal oxygen concentration. 
Although the 3d$_{x^2-y^2}$ band, the Fermi surface, and its Wannier-orbitals 
are similar for these superconductors, 
there is an apparent difference in the unoccupied levels above $E_F$. 
Based on a multi-reference density-functional-theory formulation, 
effective two-particle exchange interactions are estimated to derive 
enhancement in intra-layer exchange interactions for 
HgBa$_2$CuO$_4$, while it is weakened in TlBaLaCuO$_5$ 
and furthermore it is weak in TlBa$_2$CuO$_5$. 
The characteristic difference in the band structure is correlated 
with oxygen contents in the buffer layer. 
We also comment on the similar feature in triple-layered compounds. 
Our spin-fluctuation enhancement mechanism in an electron-correlation regime 
is consistent with the experimental fact. 
}

\begin{document}
\maketitle

\section{Introduction}

After the first discovery of a cuprate superconductor in 1986,\cite{La-first} 
plenty of cuprate superconductors have been reported.\cite{Y-first,Hg1201-first,Hg1223_1,Hg1223_2} 
Especially Hg-based cuprates are well known because of 
their high transition temperatures. 
For the high-temperature superconductivity (high $T_c$), 
in Hg-based cuprate crystal, 
the number $n$ of CuO$_2$ planes should be three 
in a basic stacked-layered structure, {\it i.e.} in the unit cell. 
The critical temperature $T_c$ for superconductivity rises up to 135K 
even at ambient pressure 
when a triple-layered Hg-based cuprate ($n=3$) is chosen 
and when it is optimized with respect to its material parameters, 
{\it e.g.} the oxygen content.\cite{Hg1223_1,Hg1223_2}
Although several other compounds have similar superconducting properties, 
$T_c$ becomes maximum by a choice of the material and 
by the adjustment of its internal material parameters. 

To explore a basic mechanism of high-$T_c$, 
we reconsider several common features and detailed difference 
among cuprates, which should be explained by a unique theory. 
One of the special characters of high-$T_c$ 
is material dependence of the superconducting properties. 
There are several known classification of cuprate superconductors in series, 
{\it e.g.} Hg-compounds, Tl-compounds, and Bi-compounds, 
which essentially differ in the atomic structure of the buffer layer. 
When we see some specific materials, and if we compare 
$T_c$ of some compounds in different series, we can actually find  
several hints to understand the relevant superconducting mechanism. 

We see global similarity in doping dependence of the CuO$_2$ planes. 
The optimal doping is often found at around a hole concentration 
of 0.16 per a CuO$_2$ plane. To adjust the concentration, 
one often needs to modify the buffer layers in its oxygen contents. 
When a high-$T_c$ material is optimized with respect to the hole concentration, 
at the optimal doping, the layer-number dependence of 
the transition temperature $T_c$ in several series of cuprates may be derived. 
Comparison among materials categorized in these series was made 
in experiment.\cite{mukuda2011high} 
Triple-layer compounds provide the highest $T_c$ among 
multi-layered compounds in various series. 
In addition, some special features of series dependence were concluded, 
{\it e.g.} existence of a large $n$ regime. 
When the layer number $n$ becomes more than around 7, $T_c$ reaches at 
a saturated value for each series. 
Actually, experimental findings of material dependence in nature 
always extend our understanding of the high-T$_c$ superconductivity. 

Theoretically, there are many successful explanations 
on tendency of the material dependence. 
An example was the Fermi surface shape dependence of 
several cuprate series.\cite{Fermishape_2,Fermishape_1} 
In this direction, Sakakibara, {\it et al.} have explored that 
the single-layer Hg-based cuprate is in a good condition for 
the orbital distillation effect,\cite{Sakakibara2010,Sakakibara2012,Sakakibara2014} 
while the 214 phase of La compounds may be a mixed multi-$d$-band system. 
They proposed that a purified 3d$_{x^2-y^2}$ band for Hg-compounds 
should provides better high-$T_c$, while hybridization of 3d$_{x^2-y^2}$ 
and 3d$_{3z^2-r^2}$ components around the Fermi level may 
causes reduction in $T_c$. Even with this understanding, 
however, there remains unresolved material dependence. 

Here an important hint can be found in the $T_c$ difference 
between the Hg-series and the Tl-series. 
A known experimental fact is that 
$T_c$ of Hg1201 is $T_c\approx 100$K and 
that of Tl1201 is $T_c\approx 50$K at the optimal doping.\cite{mukuda2011high} 
Indeed, $T_c$ of Tl1201 is only a half of the value of Hg1201. 
As we will show, the band structure calculation and the tight-binding fitting 
by the Wannierization technique\cite{wannier1,wannier2} 
tell that relevant 3d$_{x^2-y^2}$ bands 
of these compounds resemble each other. 
Conversely speaking, a single-band model derived by limited numbers 
of orbitals may not completely describe the material dependence. 

On this problem, there had been several discussions on the $T_c$ value 
and its dependence on materials parameters. 
The largeness of Madelung potential at the apical oxygen site,\cite{apex}  
the orbital energy difference between 3d$_{x^2-y^2}$ and 
3d$_{3z^2-r^2}$\cite{Sakakibara2010,Sakakibara2012,Sakakibara2014}	, the largeness of interlayer tunneling effect\cite{anderson1995interlayer}, 
the lower amount of disorder on the CuO$_2$ plane were considered. 
As for the former three factors, however, the band structure calculation 
had captured the material characteristics as far as the single-particle 
transfer terms are concerned. 
The third effect should be reconsidered with careful consideration of 
two-particle parts relevant for the interlayer pair-hopping processes 
described by an effective Hamiltonian.\cite{Nishiguchi2013}  
The last point would be experimental. However, we should note 
stiff nature of the cuprate high-temperature superconductivity 
against potential scatters.\cite{Potential_eisaki} 

Here, we propose another factor of the material dependence. 
We may discuss intra-layer effective exchange 
scattering between quasi-particles. 
The major contribution should be the super exchange mediated by 
the in-plane oxygen contributions. 
Actually, the famous Zhang-Rice mechanism of 
the singlet formation\cite{PhysRevB.37.3759} 
taught us the relevance of the in-plane exchange interaction 
along each Cu-O-Cu bond for cuprates, which was described by 
a $d$-$p$ model . 
There is, however, another exchange path mediated by elements even 
in the buffer layer. 
Owing to a renormalization formalism based on the multi-reference 
density functional theory (MR-DFT),\cite{doi:10.1143/JPSJ.70.2038} 
we can evaluate the material dependence of the exchange scattering 
among electrons in a CuO$_2$ plane. 
The formulation can start from any conventional DFT-based band structure 
calculation. By refining the representation from the single-reference 
to a multi-reference, an up-conversion Hamiltonian is defined.  
A relevant factor for our formulation is ``the super process'' inevitably 
concluded by the renormalized form 
of the strongly-correlated electron system. 

In this paper, we analyze the band structures given by 
the generalized-gradient approximation of DFT. 
By comparing the energy bands of several Hg-based and Tl-based compounds, 
we deduce a factor deciding the super process, which may promote 
the strongly-correlated superconductivity. 
The Kohn-Sham orbitals as given by the band structure calculation 
may be used to make an expanding basis for a multi-reference representation of 
the strongly-correlated system. 
As a step forward to the full construction, 
we evaluate an exchange scattering path of two electrons 
around the hot-spots, which is mediated by the orbitals given by the buffer layers. 
As a result, we can find a material dependence in 
our enhancement mechanism of the spin-fluctuation mediated superconductivity. 

\section{Methods}
\subsection{Band structure calculation}
In our MR-DFT formalism, we utilize an expression of an energy functional
\begin{eqnarray}
\lefteqn{\bar{G}_{(C)_i,\varepsilon_i}[\Psi]} \nonumber \\ 
&=&
\langle \Psi | \hat{T}+\hat{V}_{(C)_i}|\Psi \rangle 
+\frac{e^2}{2}\int d^3r d^3r' \frac{n_\Psi({\bf r})n_\Psi({\bf r}')}{|{\bf r}-{\bf r}'|}
\nonumber \\
&+&E_{\varepsilon_i}[n_\Psi]+\int d^3r v_{ext}({\bf r})n_\Psi({\bf r}). 
\end{eqnarray}
Here $|\Psi\rangle$ is a state vector representing a multi-referenced 
correlated electron state and the charge density is given by 
$n_\Psi ({\bf r}) \equiv \langle\Psi|\hat{n}({\bf r})|\Psi\rangle$. 
An exchange-correlation energy-density functional $E_{\varepsilon_i}[n_\Psi]$ 
is given in an approximated form as a model, 
so that it is differentiable with respect to the density. 
A quantum fluctuation part representing an electron-correlation effect 
in an explicit expression is given in a next form. 
\begin{eqnarray}
\lefteqn{\langle \Psi | \hat{V}_{(C)_i} | \Psi \rangle}
\nonumber \\
&=&
\langle \Psi | P_A \hat{V}_{\rm ee} P_A | \Psi \rangle
+\langle \Psi | P_A \hat{V}_{\rm ee} P_B | \Psi \rangle
+\langle \Psi | P_B \hat{V}_{\rm ee} P_A | \Psi \rangle
\nonumber \\
&-&
\int d^3r d^3r' \frac{e^2
\langle \Psi | P_A \hat{n}({\bf r}) P_A | \Psi \rangle
\langle \Psi | P_A \hat{n}({\bf r}') P_A | \Psi \rangle
}{2|{\bf r}-{\bf r}'|}
\nonumber \\
&=&
\langle \Psi | \hat{V}_{\rm ee} |\Psi \rangle
-
\langle \Psi | P_B \hat{V}_{\rm ee} P_B | \Psi \rangle
\nonumber \\
&-&
\int d^3r d^3r' \frac{e^2
\langle \Psi | P_A \hat{n}({\bf r}) P_A | \Psi \rangle
\langle \Psi | P_A \hat{n}({\bf r}') P_A | \Psi \rangle
}{2|{\bf r}-{\bf r}'|} , 
\end{eqnarray}
where projection operators $P_A$ and $P_B\equiv \hat{1} -P_A$ 
project a phase space of representation state vectors $\left\{|\Psi_A\rangle\right\}$, which is called the A-space from now on,  and its complementary space. The complementary is called the B-space. 

For this setup, we have a direct product form of 
$|\Psi_A\rangle\equiv P_A|\Psi\rangle=|\Phi_A\rangle \otimes |\Phi_0\rangle$ 
for the representation state vector and 
$|\Psi\rangle=|\Psi_A\rangle+|\Psi_B\rangle$. 
The state $|\Phi_A\rangle$ is a multi-referenced state 
with a representation in a linear combination of multi-Slater determinants. 
Elements in the combination of $|\Phi_A\rangle$ are defined 
in a selected set of bands in the Bloch representation 
(or orbitals in the Wannier representation), which are specified by 
the orbital set $A$. In a usual correlated electron system, 
the complementary $A^c$ of $A$ is given by filled core levels, 
which may includes semi-core states, and empty bands in high energy. 
The core state may be given as 
\[|\Phi_0\rangle=\prod_{l \in A^c, \varepsilon_l \le E_F} 
c^\dagger_{l\uparrow}c^\dagger_{l\downarrow}|0\rangle. \]

The definition of $A$ and the resulted set $\left\{|\Psi_A\rangle\right\}$ 
may be given by a single-particle part of 
a total effective Hamiltonian given by 
\begin{equation}
{\cal H}_{(C)_i}|\Psi\rangle = \frac{ \delta \bar{G}_{(C)_i,\varepsilon_i}[\Psi]}{\delta \langle\Psi|}.
\end{equation}
\begin{eqnarray}
{\cal H}_{(C)_i} &=& {\cal H}^1_{(C)_i} + {\cal H} ^2_{(C)_i} + {\cal H}_{(C)_i}^{1{\rm c}} ,\\
{\cal H}^1_{(C)_i} &=& \hat{T}
+\int \bar{v}_{{\rm eff},i}({\bf r})\hat{n}({\bf r})d^3r , \\
{\cal H}^{2}_{(C)_i} 
&=&
P_A \hat{V}_{\rm ee} P_A + P_A \hat{V}_{\rm ee} P_B + P_B \hat{V}_{\rm ee} P_A\\
{\cal H}^{1{\rm c}}_{(C)_i} 
&=&-\int d^3r d^3r' \frac{e^2\langle\Psi|P_A\hat{n}({\bf r}')P_A|\Psi\rangle }
{|{\bf r}-{\bf r}'|} P_A \hat{n}({\bf r}) P_A. \nonumber \\
\end{eqnarray}
Actually, the single-particle effective Hamiltonian ${\cal H}^1_{(C)_i}$ 
is given by an effective potential $\bar{v}_{\rm eff}({\bf r})$, 
\begin{equation}
\label{model-Effective-potential_1st}
\bar{v}_{{\rm eff},i}({\bf r}) = 
\int \frac{n({\bf r}')}{|{\bf r}-{\bf r}'|}d^3r' 
+ \mu_{\varepsilon_i}({\bf r})
+ v_{\rm ext}({\bf r}) ,
\end{equation}
and we have a counter part ${\cal H}^{1{\rm c}}_{(C)_i}$ defined above. 
Here, contributions in $\bar{v}_{{\rm eff},i}({\bf r})$ are 
the Hartree contribution, the model-exchange-correlation potential part, 
\begin{eqnarray}
\mu_{\varepsilon_i}({\bf r}) &=& \frac{\delta E_{\varepsilon_i}[n]}{\delta n({\bf r})},  \end{eqnarray}
and the external potential $v_{\rm ext}({\bf r})$ for the system. $v_{\rm ext}({\bf r})$ may contain a scalar potential contribution of the core electrons 
when the ionic core description is enough accurate. 
This ionic potential may be replaced by a pseudo-potential. 
Replacement of a valence orbital by the corresponding pseudo wave function may be allowed, if the quantum fluctuation contribution, requiring evaluation of the Coulomb scattering processes, is properly evaluated by a technique of the non-empirical pseudo potential method. 

By letting the number of states in $A$ larger, by reducing the model 
exchange-correlation contribution, we may make models in a series. 
When $P_A \rightarrow \hat{1}$, and when $\lim_{i\rightarrow \infty} E_{\varepsilon_i}[n] = 0$, the effective Hamiltonian approaches to the original Coulomb Hamiltonian for the electron system, and $|\Psi\rangle$ becomes the exact state of the electron system. 
Conversely, when the set of correlated orbital becomes an empty set, $A=\phi$, $|\Phi_A\rangle$ vanishes, $P_A$ projects only a single-reference state $|\Psi_A\rangle=|\Phi_0\rangle$, and our representation becomes an ordinary Kohn-Sham single-reference representation of DFT, and the effective Hamiltonian becomes the Kohn-Sham Hamiltonian without explicit two-body terms.

As a convenient approximation, we may utilize a known exchange-correlation energy functional with additional correlation term $\hat{V}_{(C)_i}$ for a small number of relevant orbitals, where the electrons show strong correlation effects. 
The charge density, when it is converged in a self-consistent calculation, 
is mainly composed of the low-energy orbitals as $|\Phi_0\rangle$ and 
an averaged contribution from $|\Phi_A\rangle$. In such a case, the counter term contribution via ${\cal H}^{1{\rm c}}_{(C)_i}$ does not have a big effect 
for the electron charge density . 
This means that a set of Kohn-Sham orbitals determined by 
${\cal H}^1_{(C)_i}$ with a converged density may give a good expanding basis 
for the MR-DFT description. The convergence may be tested by having 
difference from the reference density and the density given by 
a multi-reference state vector $|\Psi_A\rangle$. 

For cuprate, it is partially tested by having the optimized atomic structure by 
a single-reference calculation and comparing the result with the experiment. 
If the result shows good coincidence, inclusion of the low-energy correlation effect 
should not change the electron density so much from the approximated density. 
The correlation part $\hat{V}_{(C)_i}$ should only shift $|\Psi_A\rangle$ 
in a correlated state vector without big shift in the charge density. 
The momentum distribution function or the occupation of orbitals counted by $|\Psi_A\rangle$ become indicators for this test. 

In this paper, for the derivation of the single-particle effective Hamiltonian, 
we adopt the Perdew-Burke-Ernzerhof formula of 
the generalized gradient approximation\cite{PhysRevLett.77.3865} 
in DFT for the exchange correlation energy functional. 
This self-consistent scheme is known to be accurate enough to reproduce 
crystal parameters and to determine tight-binding parameters for various cupurates. 
Therefore, the inclusion of the effective two-body part should not change 
the global feature of effective occupation numbers of the Kohn-Sham orbitals 
and the charge density. 

\subsection{An effective up-conversion Hamiltonian}
Even though the structural parameters are well reproduced by conventional methods of DFT, in order to describe the strong-correlation effects for the high-$T_c$ materials, 
we need to incorporate the correlation effect into our effective model Hamiltonian. 
This is naturally done by including $\langle \Psi| \hat{V}_{(C)_i}|\Psi \rangle $ in our wave-function functional $\bar{G}_{(C)_i,\varepsilon_i}$. While the Kohn-Sham scheme is given by letting $\langle \Psi| \hat{V}_{(C)_i}|\Psi \rangle \equiv 0$. 
Therefore, we need to up-convert the ordinary Kohn-Sham Hamiltonian to a multi-referenced correlated model including $\langle \Psi| \hat{V}_{(C)_i}|\Psi \rangle $. 

In our description, we have a set of correlated bands coming from the Cu 3d$_{x^2-y^2}$ orbitals for cuprate. These bands are known to be well-defined in the DFT band structure. 
In this paper, we make a choice of a representation where only the 3d$_{x^2-y^2}$ band is relevant for the set $A$, $|\Psi_A\rangle$, and the A-space. 
The selection of the set $A$ for the 3d$_{x^2-y^2}$ orbital may be done using the Wannierization transformation. However, since our formalism is flexible, we may make use of the Bloch representation for the definition of the set $A$. 

Once we have the A-space, the representation of $|\Psi\rangle=|\Psi_A\rangle+|\Psi_B\rangle$ allows us to rewrite the determination equation ${\cal H}_{(C)_i}|\Psi\rangle = E|\Psi\rangle$ into a simplified form. Since we fix the A-space representation, we omit the symbol $(C)_i$ to represent a specified model. In addition, we may rewrite the Hamiltonian into a form of ${\cal H}=\hat{H}_{AA}+\hat{h}_B+\hat{H}_{BA}+\hat{H}_{AB}$. Here, $\hat{H}_{AA}$ is a many-body model Hamiltonian where each scattering of electrons happens only in orbitals of $A$. 
The operator $\hat{h}_B$ is a single-particle Hamiltonian describing the orbital energy in $A^c$.  
Since we subtracted $\langle\Psi|P_B\hat{V}_{\rm ee}P_B|\Psi\rangle$ in our definition of $\bar{G}_{(C)_i,\varepsilon_i}[\Psi]$, we have no direct interaction among particles in $A^c$ orbitals. Namely, the quasi-particles in $A^c$ are treated as non-interacting among them.  
The other two terms, $\hat{H}_{BA}$ and $\hat{H}_{AB}$, represent interaction processes scattering at least one particle from an orbital in $A$ to another in $A^c$ or a reverse process. 
Please note that, owing to $\hat{H}_{BA}$ and $\hat{H}_{AB}$, there appear indirect interactions among particles in $A$. 
Then, we can derive, 
\begin{eqnarray}
\lefteqn{\left(\hat{H}_{A}+\hat{h}_{B}\right) |\Psi_A\rangle} 
\nonumber \\
&+& P_A \hat{H}_{AB}\left(E - \hat{H}_{A} - \hat{h}_{B} 
- P_B \hat{H}_{AB} - \hat{H}_{BA} \right)^{-1} 
\hat{H}_{BA} P_A |\Psi_A\rangle \nonumber \\
&=& E|\Psi_A\rangle .
\label{upconversion}
\end{eqnarray}
This Brillouin-Wigner-type determination equation is exact. 
Here, the second term represents the super processes. 
\begin{equation}
\hat{H}_{\rm super} = 
P_A \hat{H}_{AB}\left(E - \hat{H}_{A} - \hat{h}_{B} 
- P_B \hat{H}_{AB} - \hat{H}_{BA} \right)^{-1} 
\hat{H}_{BA} P_A. 
\end{equation}
This expression contains a resolvent operator, which actually generates 
many-particle Green functions in a real-frequency domain. This is because $\hat{H}_{\rm super}$ operates on a multi-reference state $|\Psi_A\rangle$ and its expectation value should be given only by a many-particle Green function. 

If we note that the diagonal parts of $\hat{H}_{AA}$ and $\hat{h}_B$ do not change the number of electrons in the orbitals of $A$, and that $\hat{H}_{BA}$ and $\hat{H}_{AB}$ do change, 
we notice that the present form allows us to categorize the many-particle Green functions. Namely, when the resolvent in $\hat{H}_{\rm super}$ is counted, the Green function coming out of this expression should have a non-trivial quasi-particle number at least in one of an electron quasi particle set or a hole quasi particle set in $A^c$. 

\subsection{Two-particle scattering amplitudes} 
The last up-conversion Hamiltonian $\hat{H}_{A}+\hat{h}_{B}+\hat{H}_{\rm super}$ 
defined by Eq.~(\ref{upconversion}) is formally exact when we treat $\bar{G}_{(C)_i,\varepsilon_i}[\Psi]$ without approximation. 
The first term $\hat{H}_{A}$ contains direct interactions among electrons in the correlated $A$ orbitals. While, the super process $\hat{H}_{\rm super}$ represent indirect interactions for these correlated electrons. 
We can readily show that the screening effects owing to the electron-electron interaction are derived from $\hat{H}_{\rm super}$. The screened effective interaction is given by summing the direct interaction of $P_A\hat{V}_{ee}P_A$ and $\hat{H}_{\rm super}$. If we do a partial summation of the bubble diagrams appearing in $\hat{H}_{\rm super}$, we may easily derive the RPA screening. The formally exact form of $\hat{H}_{\rm super}$ contains every scattering diagrams in the higher order. Actually, the ladder diagram can be readily derived from the expression of $\hat{H}_{\rm super}$, and the vertex corrections are systematically derived. 

In this paper, however, we focus on material dependence, which may be well described by our $\hat{H}_{\rm super}$. Therefore, we treat $\hat{H}_{\rm super}$ in an approximated manner. This expression contains final-stete effects coming from the exact resolvent part. It requires a precise evaluation of high-energy steady states of the model system. However, for the analysis of cuprates, we may use a characteristic band structure, where the 3d$_{x^2-y^2}$ band has a partial energy window. If the low-energy phenomena relevant for high-$T_c$ happens essentially in a correlation effect within this energy range, we may treat material dependence of the effective low-energy scattering processes by utilizing a gap between the other bands. Namely, below the energy window around $E_F$ of the DFT-GGA band, we have oxygen bands. Above the window, we have several high energy bands. Owing to the energy separation, we can safely evaluate a contribution from $\hat{H}_{\rm super}$ in an approximated manner by making a simple approximation for the resolvent. 

In the exchange scattering, or the exchange interaction, it is well-known that the screening effect may not be relevant in general. In our evaluation of $\hat{H}_{\rm super}$, due to the above-mentioned character of the band scheme, we may be allowed to evaluate the material dependence in the exchange scattering by neglecting the higher order screening effects. Actually, the screening of 
intra-band direct exchange process does not have relevant contribution compared to the indirect exchange process discussed in Sec.~\ref{single-layer}. 

\section{Results}
\subsection{Hg- and Tl-based cuprate compounds}

For comparison, we consider Hg-compounds and Tl-compounds. 
When a cuprate crystal is prepared at an optimal doping, the structure 
often becomes an alloy or a mixed phase. 
To adopt a reliable DFT code\cite{qe} in our simulation, however, 
we need to consider a perfect periodic crystal with a unit cell. 
Therefore, the simulation becomes possible by limiting 
the filling factor at some special value allowing construction of a super cell. 
Owing to this reason, we treat a crystal phase fixing concentration of 
dopant and an oxygen composition ratio. 
In some cases, we look at a filling factor corresponding to 
the half-filling of the CuO$_2$ plane. Obtaining an expanding basis via DFT-GGA, 
our MR-DFT derives an effective many-body Hamiltonian, whose physical phase 
becomes a Mott insulating state for these cases. 
Note that the obtained band structure merely 
gives a spectrum of a single-particle part of the many-body effective Hamiltonian. 
For a good Mott insulator, therefore, the band structure of this single-particle 
part should show metallic features rather than an insulating gapped phase. 

\begin{figure}[htbp]
 \begin{minipage}[b]{0.4\linewidth}
  \centering
 \hspace*{+2em} 	
 \includegraphics[keepaspectratio, scale=0.28]{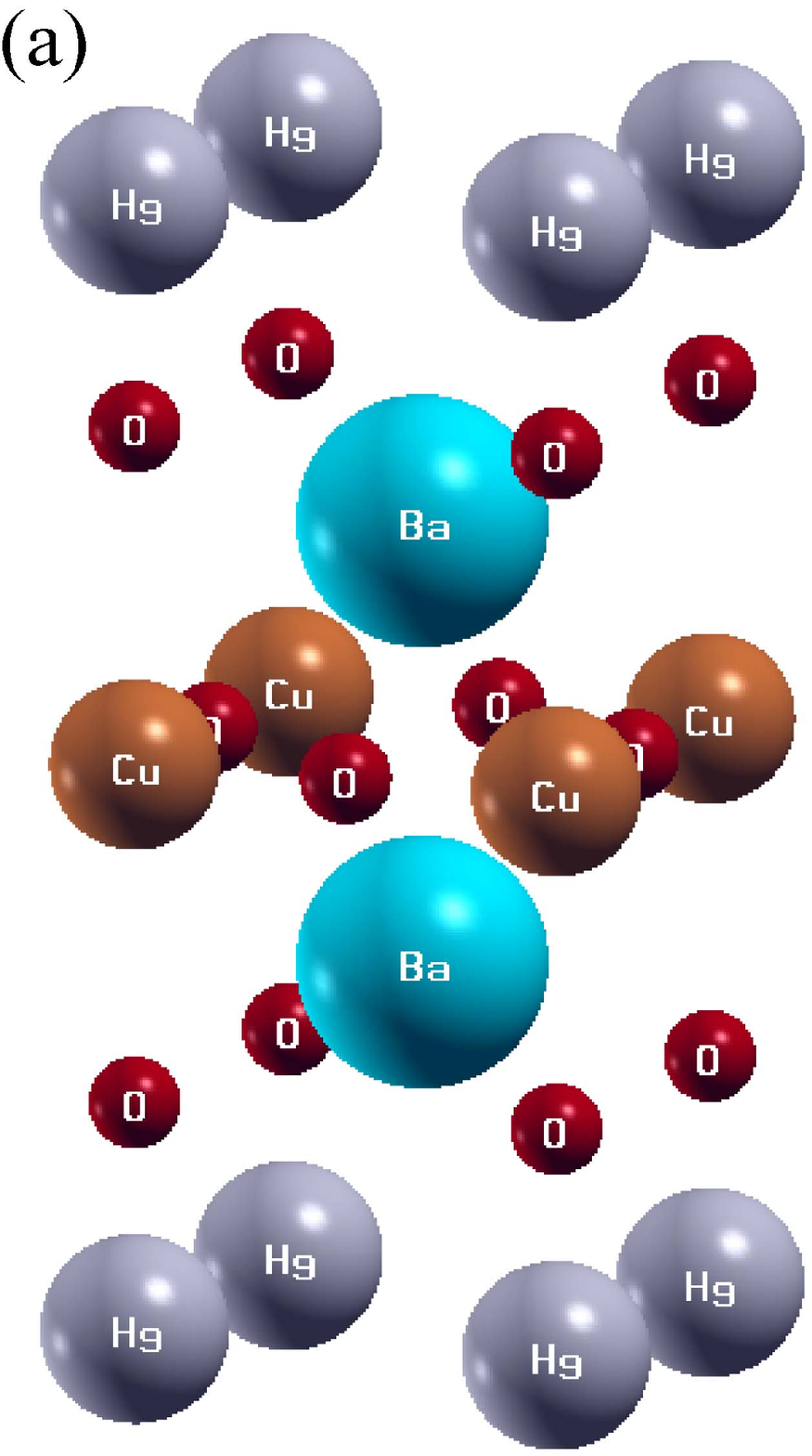}
 \end{minipage}
 \vspace*{2em}
 \begin{minipage}[b]{0.4\linewidth}
  \centering
 \hspace*{2em} 
  \includegraphics[keepaspectratio, scale=0.26]{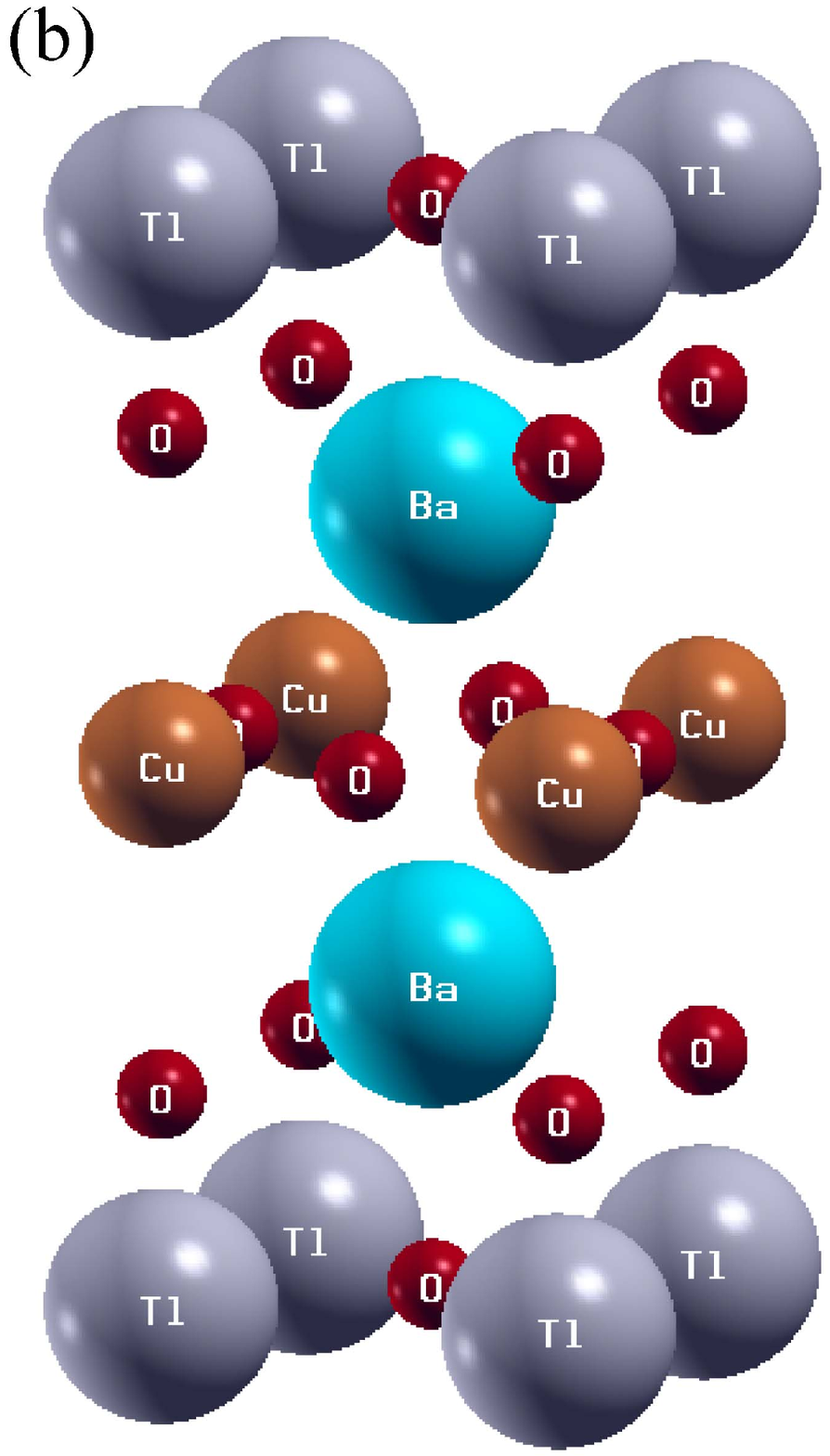}
 \end{minipage}
\caption{The atomic structures of (a) HgBa$_2$CuO$_4$(left) and  
(b)TlBa$_2$CuO$_5$(right). \label{fig:structure}}
\end{figure}

For the filling control, we mainly consider creation of oxygen deficiency. 
The oxygen concentration is controlled 
by reduction/oxidation of the buffer layers. 
Therefore, when we consider, for example, 
HgBa$_2$CuO$_{4+\delta}$ with $\delta$ from 0 to 1, 
we treat oxygen deficiency at the HgO$_\delta$ plane maintaining 
perfectness of CuO$_2$ layers. 
Similarly, when we consider three-layer $n=3$ compounds, 
we consider the oxygen deficiency in the buffer layers. 
For the single-layer structures, we treat 
HgBa$_2$CuO$_4$ and TlBa$_2$CuO$_5$. 
The structural parameters for each crystal are determined by 
the optimization simulation in DFT-GGA, 
where the external pressure condition is zero. 
(Fig.~\ref{fig:structure}) 

HgBa$_2$CuO$_4$ lacks oxygen atoms at each HgO$_\delta$ plane. 
(Fig.~\ref{fig:structure} (a)) 
We have a local OHgO structure along the $c$ axis.  
Oxygen atoms in this OHgO structure may be 
interpreted as apical oxygen atoms of CuO$_4$ pyramids. 
While, TlBa$_2$CuO$_5$ has TlO planes. (Fig.~\ref{fig:structure} (b)) 

A nominal filling factor in the CuO$_2$ plane 
may be counted by the rule for ionization valence of 
noble metal, alkaline earth metal ions, and oxygen. 
Supposing Hg$^{+2}$, Tl$^{+3}$, Ba$^{+2}$, and O$^{-2}$, 
the Cu formal valence for these compounds are 
+2 (half-filling) for HgBa$_2$CuO$_4$ and 
+3 (one hole doped) for TlBa$_2$CuO$_5$. 
For Tl-compounds, we also examine TlBaLaCuO$_5$ 
in the discussion section, Sec.~\ref{discussion_section}
where the Cu valence becomes +2 (half-filling). 

As for the multi-layered cuprates, we focus on $n=3$ in the following sections. 
Among the three layered compounds, we will consider Hg1223 and Tl1223 
structures. Specifically, we treat HgBa$_2$Ca$_2$Cu$_3$O$_8$ 
and TlBa$_2$Ca$_2$Cu$_3$O$_9$. 
Their Cu valences are +2 (half-filling) and +7/3 (1/3 hole doping). 
Band structures of these compounds will be compared to derive 
a characteristic feature of the multi-layer compounds from our point of view. 

\subsection{Single-layer cuprate superconductors}
\label{single-layer}
\begin{figure}[htbp]
 \begin{minipage}[b]{0.32\linewidth}
 \hspace*{-8em} 	
  \includegraphics[keepaspectratio, scale=0.8]{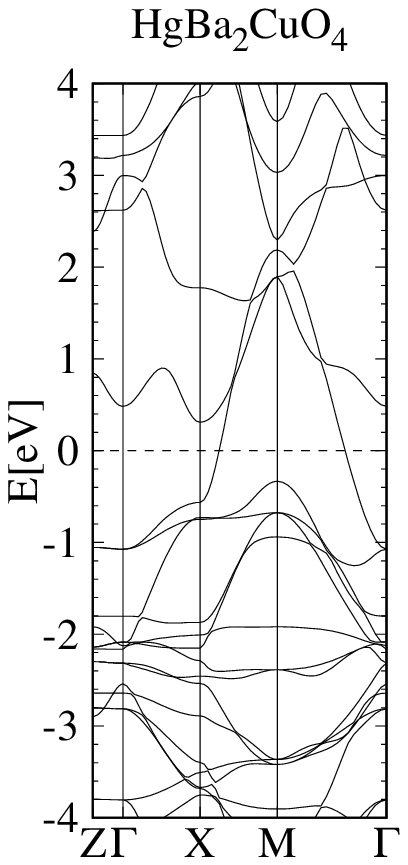}
 \end{minipage}
 \begin{minipage}[b]{0.32\linewidth}
 \hspace*{-5em} 
  \includegraphics[keepaspectratio, scale=0.8]{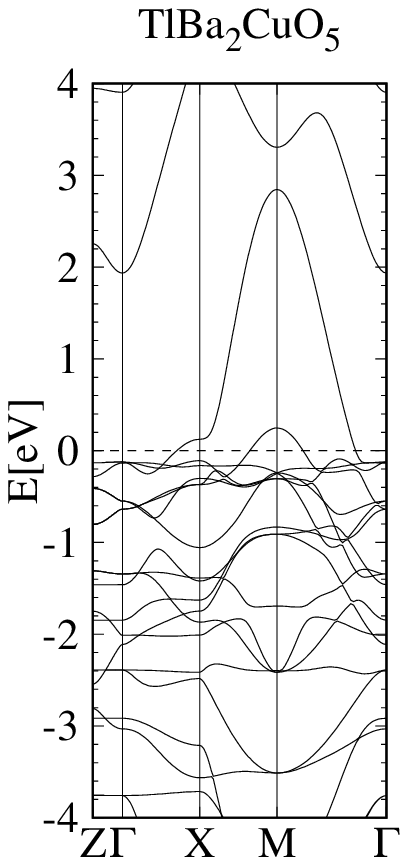}
 \end{minipage}
\caption{Band structures of HgBa$_2$CuO$_4$ (left) and TlBa$_2$CuO$_5$ (right).\label{fig:band_single}}
\end{figure}

We show the band structures of HgBa$_2$CuO$_4$, and  
TlBa$_2$CuO$_5$ in Fig.~\ref{fig:band_single}.
In this calculation, we utilize the norm-conserving pseudo potentials 
with the PBE functional. The energy cut-offs in the plane-wave expansion for 
the wave function and the charge density are (100, 400) Ry. 
For the self-consistent charge density construction, 
the integration with respect to the k vectors is done 
using a 8$\times$8$\times$8 k-point mesh in the 1st Brillouin zone. 
The unit cell of these compounds are optimized in the simulation, 
where the pressure control is done with 
a criterion of each diagonal element of the stress tensor being less than 0.5kbar. 
The internal atomic structures 
are optimized with a criterion that the summation of the absolute values 
of force vector elements becomes smaller than $1.0\times 10^{-8}$[Ry/a.u.]. 
We adopted the wannierization method with the projection operators 
for the disentanglement process.\cite{wannier1,wannier2} 

The major parameters of the effective hopping Hamiltonian 
(the tight-binding model) for the 3d$_{x^2-y^2}$ band of HgBa$_2$CuO$_4$ 
are similar to former estimations in the literatures.\cite{Sakakibara2010} 
In addition, the values of $t$, $t'$, and $t''$ are nearly the same for 
these compounds. (Table~\ref{table:hopping})
\begin{table}[hbtp]
  \caption{The parameters of the effective hopping hamiltonian for the 3d$_{x^2-y^2}$ band in Hg1201,Tl1201 and Tl2201.
  \label{table:hopping}}
  \centering
  \begin{tabular}{lccc}
    \hline
& Hg1201  &  Tl1201  \\
    \hline \hline
  $t$  &  -0.452 &  -0.574 \\
    $t'$  &   0.112 & 0.0990 \\
   $t''$  &  -0.100 & -0.0823 \\
    \hline
  \end{tabular}
\end{table} 

Here, a little larger value of $t$ for the Tl-compound comes 
from its shrunk lattice constant. When we use the lattice constant 
of Tl1201 for Hg1201, the value of $t$ becomes close to -0.56. 
Considering change in $T_c$ in high pressure for Hg-compound, 
the difference in the transition temperature can not be fully explained by 
the in-plane transfer parameters. 
Except when the effective screened interaction 
$U$ is strongly material dependent, it is not easy to derive 
the reduction of $T_c$ for Tl1201 
by modeling with a single-band Hubbard Hamiltonian. 
We find that the oxygen bands coming from CuO$_2$ plane 
are also similar for these compounds. 
This is rather natural, because difference in the material structures 
among Hg1201 and Tl1201 compounds 
comes essentially from the buffer layers. The difference in the in-plane 
lattice constant is not enough large to modify the quantitative values 
of intra-layer hopping parameters. 

When we compare these band structures, an apparent difference 
is found in the low-lying unoccupied bands around $E_F$. 
For the HgBa$_2$CuO$_4$, it is the branch coming down 
to $E_F$ around the $X$ point and the $\Gamma$ point. 
This band originates from the Hg 5p band. 
The partial density of states explicitly shows this character. 
The level is a hybridized orbital made from Hg 5p and apical oxygen 2p levels. 
Contrary to HgBa$_2$CuO$_4$, in the band structure of Tl1201, we have no signal of 
the low energy band at the $X$ point. The Cu 3d$_{x^2-y^2}$ band of 
TlBa$_2$CuO$_5$ is well isolated from the upper branches showing 
a good distillation in this sense. The specialty of TlBa$_2$CuO$_5$ becomes further 
explicit by comparing this heavily hole-doped phase to La-doped Tl1201. 
In a crystal phase of TlBaLaCuO$_5$, interestingly, 
we see the branch of TlO coming down to $E_F$ around the $\Gamma$point.\cite{TlBaLaCuO5}
Therefore, this characteristic feature coming from the band structure 
has a good correlation with $T_c$ of these compounds. 

\begin{figure}[h]
\begin{center}
\includegraphics[width=5.0cm]{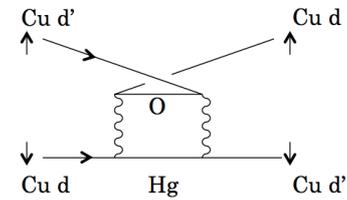}
\caption{Spin exchange process via low-lying quasi-particle (Hg) and quasi-hole (O) levels inducing enhanced electron correlation in Hg compounds. \label{fig:exchange}}
\end{center}
\end{figure}

In our simulation method, we evaluated the exchange scattering amplitude 
of a process, 
\begin{equation}
H_{\rm scat} = 
\sum_{{\bf k},{\bf k}',\sigma,\sigma'} J^{\rm eff}_{{\bf k},{\bf k}'} 
c^\dagger_{{\bf k}',\sigma} c^\dagger_{{\bf k},\sigma'} 
c_{{\bf k}',\sigma'} c_{{\bf k},\sigma}. 
\end{equation}
Here, the amplitude is explicitly given by 
\begin{equation}
J^{\rm eff}_{{\bf k},{\bf k}'} 
= - \sum_{{\bf K},n\in A^c_1,m\in A^c_2} 
\frac{(V_{\rm ee})_{ {\bf k},(m,{\bf K}):(n,{\bf K}),{\bf k} }(V_{\rm ee})_{(n,{\bf K}),{\bf k}':{\bf k}'(m,{\bf K})}}{\varepsilon_{{\bf K},n}+\tilde{\varepsilon}_{{\bf K},m}}. 
\end{equation}
In the summation, $n\in A^c_1$ represents higher empty bands 
and $m\in A^c_2$ represents filled bands for holes. 
The amplitude of $(V_{ee})_{(n,{\bf K}),{\bf k}':{\bf k}'(m,{\bf K})}$ is given by,
\begin{eqnarray}
\lefteqn{(V_{\rm ee})_{(n,{\bf K}),{\bf k}':{\bf k}'(m,{\bf K})} } \nonumber \\
&=&\int d^3r d^3r' \frac{e^2 \phi_{{\bf k}'}^*({\bf r}) \varphi_{n,{\bf K}}^*({\bf r}')
\phi_{{\bf k}'}({\bf r}')\varphi_{m,{\bf K}}({\bf r}) }{2|{\bf r}-{\bf r}'|}. \end{eqnarray}
Two types of Kohn-Sham orbitals are $\phi_{{\bf k}'}({\bf r}')$ for the correlated 3d$_{x^2-y^2}$ band and $\varphi_{m,{\bf K}}({\bf r})$ for the other levels in $A^c$. 
The process is schematically described by a Feynman diagram of Fig.~\ref{fig:exchange}. 
In the evaluation of $J^{\rm eff}$, we used 4$\times$4$\times$4 k-mesh points 
with the cut-off energy of (40,160) Ry to save the computation time. 

Since this process for {\it e.g.} a Hg compound is 
mediated by the Hg-band and the oxygen 2p band, 
it is not derived from a $d$-$p$ model. 
Interestingly, the enhanced $J^{\rm eff}$ of O(10)meV is 
concluded for Hg1201 around the $X$ point 
{\it i.e.} the hot spot. (Table~\ref{table:Jeff})
However, for Tl1201, the value of $J^{\rm eff}$ is O(1)meV much smaller than 
the others, so that it would be negligible. 
When we adopt the spin-fluctuation mechanism in 
a correlated electron model, the enhanced exchange 
naturally have a positive effect to enhance $T_c$. 

\begin{table}[hbtp]
  \caption{Evaluated amplitudes for the exchange process in Hg1201, Tl1201 
and La-doped Tl1201. The values $J^{\rm eff}_{X,Y}$ are 
$J^{\rm eff}_{{\bf k},{\bf k}'}$ in eV when ${\bf k}$ and ${\bf k}'$ are at 
the $X$ point and $Y$ point, respectively.   \label{table:Jeff}}
  \centering
  \begin{tabular}{lccc}
    \hline
& HgBa$_2$CuO$_4$  & TlBa$_2$CuO$_5$ 
& TlBaLaCuO$_5$ \\
    \hline \hline
  $J^{\rm eff}_{X,Y}$  
& -0.04  &  0.004 
& -0.008 \\
    \hline
  \end{tabular}
\end{table} 

To show the relevance of the buffer layer structure, 
we mention a result for TlBa$_2$CuO$_4$. 
In this imaginative structure, we have the low-energy branch 
down even below $E_F$. 
Therefore, the reduction of oxygen contents in the buffer layer 
is rather important for this mechanism of enhancement in $J^{\rm eff}$. 
Unfortunately, for the single-layer Tl compound, 
the mechanism seems not to be apparent for a reasonable 
value of oxygen contents for the superconductivity. 
This example rules the importance of oxygen atoms at the TlO layer, 
not only the apical oxygen and its geometrical parameter dependence. 

Here we should note that $J^{\rm eff}_{{\bf k},{\bf k}'}$ 
are in the momentum representation. 
There is no explicit $z$-component dependence on this value. 
So that the interaction is purely two dimensional. 
In the CuO$_2$ plane, the scattering coming from $U$ has an amplitude $U/N$ 
with the total number of mesh points in two-dimensional Brillouin zone. 
If we compare $U/N$ with $N=16$ to $J^{\rm eff}$, we can say that 
the effect of the buffer layer to enhance the spin exchange is non negligible for some materials. There exists material dependent contribution for 
the enhancement as suggested by Table~\ref{table:Jeff}. 

If we consider a screening effect of the on-site Hubbard $U$, 
a larger reduction in the value of $U$ should be expected 
when the higher (and lower) bands other than 
the correlated 3d$_{x^2-y^2}$ band are take part in the screening process. 
Actually, our scheme suggests even lowered Hubbard $U$ rather for 
HgBa$_2$CuO$_4$ than TlBa$_2$CuO$_5$. 
Therefore, the value of $U$ may not be enough for the explanation of 
material dependence even if a modeling with the single band model is adopted. 
We expect that a relatively large $U$ scheme with enhanced 
$J_{\rm eff}$ mediating the stronger spin fluctuation would be a solution for the explanation of known material dependence in $T_c$. 

\subsection{Triple-layer cuprate superconductors}
\begin{figure}[htbp]
 \begin{minipage}[b]{0.32\linewidth}
 \hspace*{-6em} 	
  \includegraphics[keepaspectratio, scale=0.6]{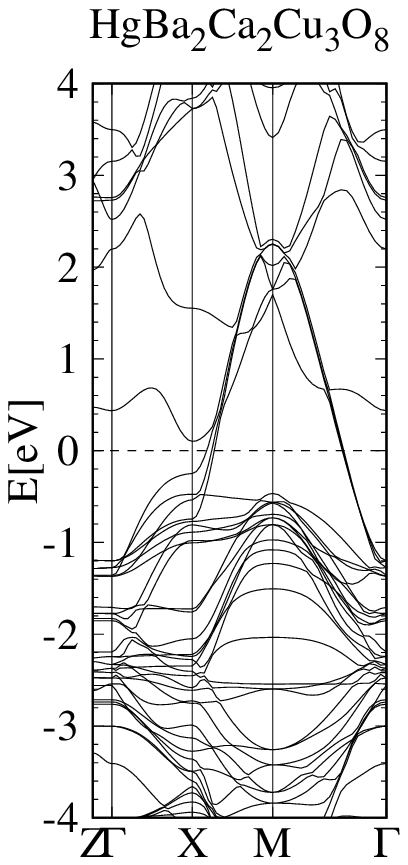}
 \end{minipage}
 \begin{minipage}[b]{0.32\linewidth}
  \hspace*{-6em} 
  \includegraphics[keepaspectratio, scale=0.6]{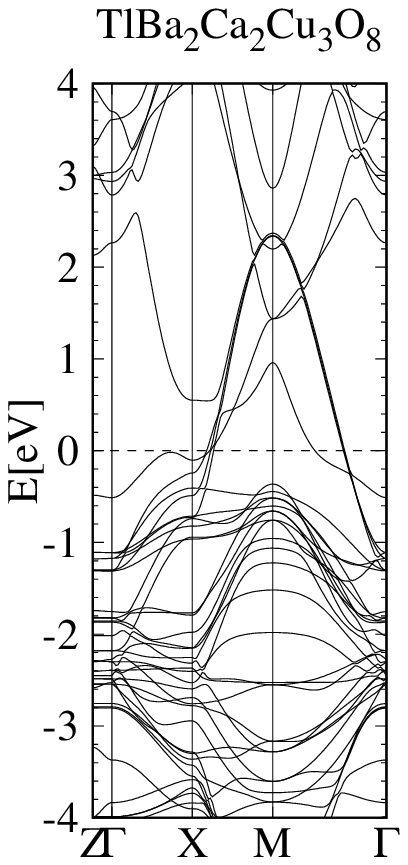}
 \end{minipage}
 \begin{minipage}[b]{0.32\linewidth}
  \hspace*{-6em} 
  \includegraphics[keepaspectratio, scale=0.6]{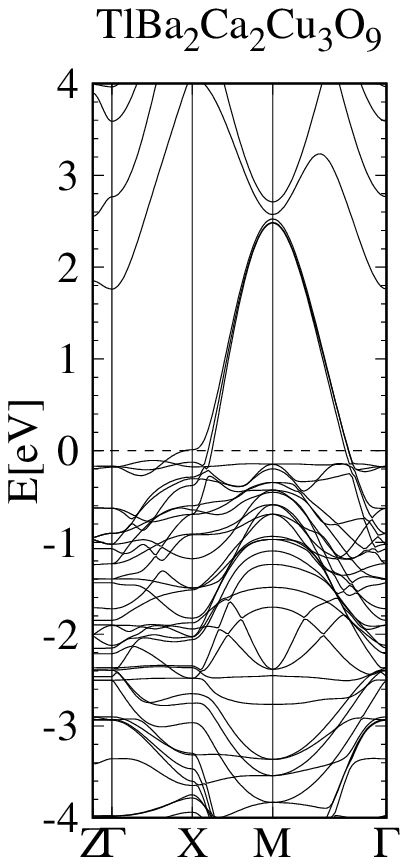}
 \end{minipage}
 \vspace{+3em}	
 \caption{Band structures of HgBa$_2$Ca$_2$Cu$_3$O$_8$ (left), TlBa$_2$Ca$_2$Cu$_3$O$_8$ (middle), and 
TlBa$_2$Ca$_2$Cu$_3$O$_9$ (right).  \label{fig:band_three}}
\end{figure}

In order to see that the mechanism derived for the single-layered cuprate in the last section 
may hold for the other multi-layer cuprates, 
we discuss the triple-layer phases of Hg- and Tl-compounds. 
In Fig.~\ref{fig:band_three}, we show the electronic band structure 
given by DFT-GGA simulations for HgBa$_2$Ca$_2$Cu$_3$O$_8$, TlBa$_2$Ca$_2$Cu$_3$O$_8$ and TlBa$_2$Ca$_2$Cu$_3$O$_9$. 

In the Hg-based triple-layer compound, we see the Hg-O branch 
coming down at the $X$ point and the $\Gamma$ point close to $E_F$. 
Therefore, the enhancement of $J_{\rm eff}$ is expected. 
Since the process is mediated by the Hg-O band, its effect should be 
bigger for the outer CuO$_2$ plan of this stacked CuO$_2$ layers. 
Experimentally, it is known that the superconductivity is much more stabilized for 
the outer plane than the inner plane. 
Our mechanism does not contradict to this known fact. 

For the Tl-compounds, we use two typical crystal structures. 
The band structure of TlBa$_2$Ca$_2$Cu$_3$O$_9$ with 1/3 hole doping 
shows similar nature to TlBa$_2$CuO$_5$, where 
there are no higher branch close to $E_F$. We cannot expect 
the enhancement mechanism for this Tl compound. 
When we see an imaginative TlBa$_2$Ca$_2$Cu$_3$O$_8$, 
we see a branch of Tl-O coming down even below $E_F$. 
These two simulations suggest that the reduction in Tl-O layers 
may cause the Tl-O band downward. 
Suppose that the branch is just at a reasonable range 
when we adjust the oxygen content to make the optimal hole concentration,\cite{hase2001carrier,liu1995superconductivity}
our enhancement mechanism may contribute the high-T$_c$ of 
the triple-layer phase of the Tl compound similar to Hg compounds. 
Therefore, as for our enhancement mechanism, 
it works nicely for Hg compounds when it is optimized for the hole concentration. 
However, it may not work well when Tl compound is considered.
\section{Discussion} 
\label{discussion_section}

In the last section, we derived material dependence 
among Hg- and Tl-compounds in the low-energy empty levels 
above $E_F$ and the resulted enhancement in effective exchange scatterings. 
This mechanism to promote the spin-fluctuation 
mediated superconductivity is supported by the levels going down 
around the $X$ point of the mercury compounds. 
One might suspect that the band effect may be originated 
by a special buffer layer with reduced amount of oxygen 
assumed for the simulation. 
 
To assess the validity of our explanation for the real materials, 
we tested some material structures. 
One is HgBa$_2$CuO$_5$ to consider the real structure of hole-doped Hg1201. 
Since HgBa$_2$CuO$_5$ is an imaginative structure prepared for the test, 
the Cu valence becomes Cu$^{+4}$ 
and the nominal hole concentration becomes two per Cu. 
The band structure shows no more low-lying branch at the $X$ point. 
In that sense, our mechanism comes from the reduced buffer layers 
with small amount of oxygen. 
In the Hg compounds, however, the stoichiometry of 
HgBa$_2$CuO$_4$ corresponds to the half-filling. 
The global band structure is usually not so much 
affected by the inclusion of slight oxygen contents in the buffer layer. 
So, our mechanism and effective Hamiltonian derived using the end material 
should naturally work for the mercury compounds. 

In Tl-compounds, it is not so easy to access 
the optimal doping by the reduction of the buffer layer. 
In a real Tl superconductor, La substitution was used to 
adjust the filling of CuO$_2$. 
We tested TlBaLaCuO$_5$, where the formal valence of Cu becomes +2. 
The result of the band structure calculation tells us that 
no signal of the low-energy branch at the $X$ point. 
In the real material of TlBa$_{1-x}$La$_x$CuO$_5$, 
it is known that $T_c$ does not reach the value over 50K even by adjusting the hole concentration.\cite{Tl1201,subramanian1990tlba1}
Thus, we may propose to consider a careful control of oxygen 
in Tl1201 without La doping, which might make an enhancement of $T_c$. 
Actually, this way of approach is consistent with some reported facts.\cite{subramanian1994stabilization}

\section{Summary and conclusions}

In this paper, we introduced a simulation method of 
the exchange scattering amplitudes in cuprates 
based on our MR-DFT method. 
By applying the method, we evaluated an effective exchange 
scattering amplitude for some Hg and Tl compounds. 
Analyzing the nature of the electronic band structures of 
these compounds, we found that the high energy levels 
originated from the Hg-O buffer layer contributes well 
creating enhancement of $J_{\rm eff}$, which should lead 
the increase in $T_c$ via the spin fluctuation mechanism. 
In Tl compounds, however, the enhancement is not so apparent. 
This is because of the absence of the Tl-O branch around $E_F$. 
Our mechanism, which is consistent with the known experimental facts, 
may explain strange difference between $T_c$s of Hg- and Tl-compounds, 
thereby it may lead us to further approaches to create the new design of 
cuprate superconductors.

\begin{acknowledgment}
The calculations were done in the computer centers of Kyushu University 
and ISSP, University of Tokyo. 
The work is supported by joint-project for 
``Study of a simulation program for the 
correlated electron systems'' with Advance-soft co. J161101009, 
and JSPS KAKENHI Grant Numbers JP26400357. 
\end{acknowledgment}


\bibliographystyle{jpsj}
\bibliography{Cuprate_super}

\end{document}